\documentclass[a4paper,fleqn,usenatbib]{mnras}

\usepackage{color}
\usepackage{graphicx}
\usepackage{amsmath}
\usepackage{amssymb}
\usepackage[T1]{fontenc}
\usepackage{mathptmx}
\usepackage{txfonts}
\usepackage{ae,aecompl} 

\title[Estimating the distribution of rest-frame timescales for blazar jets.]{Estimating the distribution of rest-frame timescales for blazar jets: a statistical approach.}
\author[Liodakis et al.]
{I.Liodakis$^{1,2}$\thanks{liodakis@physics.uoc.gr}, D. Blinov$^{1,2,3}$, I. Papadakis$^{1,2}$ and V. Pavlidou$^{1,2}$\\
$^{1}$Department of Physics and ITCP\thanks{Institute for Theoretical
  and Computational Physics, formerly Institute for Plasma Physics}, University of Crete, 71003, Heraklion, Greece\\
$^{2}$Foundation for Research and Technology - Hellas, IESL, Voutes, 7110 Heraklion, Greece\\
$^{3}$Astronomical Institute, St. Petersburg State University,Universitetsky pr. 28, Petrodvoretz, 198504 St. Petersburg, Russia\\
}

\begin{document}
\maketitle
\label{firstpage}

\begin{abstract}
In any flux-density limited sample of blazars, the distribution of the timescale modulation factor $\Delta t'/\Delta t$,  which quantifies the change in observed timescales compared to the rest-frame ones due to redshift and relativistic compression follows an exponential distribution with a mean depending on the flux limit of the sample. In this work we produce the mathematical formalism  that allows us to use this information in order to uncover the underlining rest-frame probability density function of measurable timescales of blazar jets. We extensively test our proposed methodology using a simulated FSRQ population with a 1.5 Jy flux-density limit in the simple case (where all blazars share the same intrinsic timescale), in order to identify limits of applicability and potential biases due to observational systematics and sample selection. We find that for monitoring with time intervals between observations longer than $\sim$30\% of the intrinsic timescale under investigation the method loses its ability to produce robust results. For time intervals of $\sim$3\% of the intrinsic timescale the error of the method is as low as 1\% in recovering the intrinsic rest-frame timescale. We applied our method to rotations of the optical polarization angle of blazars observed by RoboPol. We found that the intrinsic timescales of the longest-duration rotation event in each blazar follows a narrow distribution, well-described by a normal distribution with mean 87 days and standard deviation 5 days. We discuss possible interpretations of this result.
\end{abstract}

\begin{keywords}
galaxies: active -- galaxies: jets -- galaxies: blazars 
\end{keywords}

\section{Introduction}\label{intr}

Blazars are among the most active of galaxies. Their jets are known to show unique properties, such as superluminal motion, boosted emission, and high variability throughout the entire electromagnetic spectrum from $\gamma$-rays to radio. These unique properties are attributed to the preferential alignment of their jet \citep{Readhead1978,Blandford1979,Scheuer1979,Readhead1980}. Due to beaming, small differences in the rest frame result in large scatter of observables, hampering phenomenological studies and dispersing correlations and behavior trends. This also constitutes the reason why there are still many open questions regarding blazars, despite many years of systematic study.  Even today very little is known about blazar properties in their rest frame. In this work we focus on studies of blazar jets in the time domain.

Timescales in blazars are compressed due to Doppler boosting, but are also elongated due to the expansion of the Universe. This distortion is quantifiable by the timescale modulation factor $m$, 
\begin{equation}
m=\frac{\Delta t'}{\Delta t}=\frac{1+z}{D},
\label{tscale_mod}
\end{equation}     
where $\Delta t'$ is the timescale in the observer's frame, $\Delta t$ is the timescale in the rest frame of the jet, $z$ is the redshift and $D=\sqrt{1-\beta^2}/(1-\beta\cos\theta_j)$ is the Doppler factor, where $\theta_j$ is the angle to the line of sight, and $\beta$ the velocity of the jet in units of speed of light. Even though redshift is, in most cases, a well measured quantity, measuring the amount of boosting in a jet is not straight forward \citep{Liodakis2015-II}. Although several methods have been proposed (e.g. \citealp{Ghisellini1993,Readhead1994,Lahteenmaki1999-III,Liodakis2016}), the amount of available data in the literature is limited, making a more broad statistical study of timescales unfeasible. In addition, current Doppler factor estimates involve large uncertainties \citep{Liodakis2015-II}, so ``correcting'' the measured timescales on a source-by-source basis using available Doppler factor estimates may in fact introduce even more dispersion in the distribution of observables. In order to overcome such limitations, we are proposing a methodology that can estimate the effect of boosting at a population level by using the Doppler factor distribution in flux-density limited samples predicted by blazar population models.

Information on rest-frame characteristic timescales can provide important information on blazar jets and their emission processes. For example rest-frame time delays between frequencies are indicative of the distance between different emission regions \citep{Fuhrmann2014,Max-Moerbeck2014}. Rest-frame durations of flares can test different models of the evolution of jet disturbances \citep{Hovatta2007,Hovatta2008,Hovatta2009}. Rest-frame durations of EVPA rotations can help us distinguish between mechanisms proposed to interpret them \citep{Blinov2015,Blinov2016,Blinov2016-II}.

This paper is organized as follows. In \S \ref{model} we describe our model for the blazar population. In \S \ref{intr_prob_den_func} we present our mathematical formalism for uncovering the probability density function of rest-frame timescales. In \S \ref{method_app} we benchmark our method with respect to the time interval between observations and sample size. In \S \ref{robopol_int_time} we apply our method to timescales associated with EVPA rotations as seen by the RoboPol survey as well as possible interpretations of the results, and in \S \ref{sum & concl} the conclusions derived from this work. 

The cosmology we have adopted throughout this work is $H_0=71$ ${\rm km \, s^{-1} \, Mpc^{-1}}$, $\Omega_m=0.27$ and $\Omega_\Lambda=1-\Omega_m$ \citep{Komatsu2009}.

\section{Blazar population Model}\label{model}

Although blazars exhibit very diverse behavior on a source-by-source basis, they can be statistically treated as a population of relativistically boosted sources with relatively simple (power-law) distributions of rest-frame properties (luminosities and Lorentz factors) viewed at random angles. Typically such population models (see e.g \citealp{Urry1991,Padovanni1992,Vermeulen1994,Lister1997}) are developed to fit the observed luminosity and flux-density distributions.

However, because blazars are known to show high variability throughout the electromagnetic spectrum, single-epoch flux densities can be unreliable observables. For this reason, in a recent work we used observables that are not as affected by variability: the observed apparent velocity and redshift distributions of the MOJAVE \footnote{Monitoring Of Jets in Active galactic nuclei with VLBA Experiments, http://www.physics.purdue.edu/MOJAVE/} sample \citep{Lister2005,Lister2009,Lister2013}. MOJAVE uses a statistically complete flux-density limited sample, selected at 15 GHz \citep{Lister2005}, ideal for population studies of blazars. Below we present a summary of the our population models  (\citealp{Liodakis2015}, hereafter Paper~I, \citealp{Liodakis2016-II}, hereafter Paper~II). For a more detailed description of the model, results and applications see Paper~I,  \cite{Liodakis2015-II}, and Paper~II.

We assumed single power-law distributions for the Lorentz factor and the unbeamed luminosity for the simulated parent population. We treated BL Lacs and the FSRQs as separate cases in order to assess the difference in beaming between the two classes. We adopted pure luminosity evolution \citep{Padovanni1992}, where sources become brighter with look-back time while maintaining a constant comoving volume density, making the number of sources in any redshift bin proportional to the comoving volume element in that bin ($dN\propto dV$). This scaling is normalized so that the final simulated sample for the comparison with the observed distributions will consist of $ \sim 10^3$ sources. We used a Monte-Carlo approach to calculate the simulated distributions of the observables we consider: we drew values for the intrinsic luminosity $L_v$ and Lorentz factor $\Gamma$ from power-law distributions, and for the viewing angle $\theta_j$ from a uniform distribution. We then calculated a flux density using,
\begin{equation}\label{flux_dens}
S_\nu=\frac{L_\nu D^p}{4\pi{d_L^2}}(1+z)^{1+s},
\end{equation} 
where $D=1/\Gamma(1-\beta\cos\theta)$, $\beta=(1-\Gamma^{-2})^{1/2}$, $d_L$ is the luminosity distance, $s$ the spectral index, and $p$ is defined as $p=2-s$; and we finally applied the 1.5 Jy flux-density limit of the MOJAVE sample.

Following the procedure described above, we produced simulated samples for the blazar populations (Paper~I), from which we extracted information regarding the Doppler factor distribution, the distribution of $\Gamma\theta$ which quantifies how beamed is a source within a flux-density limited sample, and the distribution of the timescale modulation factor in each class.

In the original version of the model, we used a single value for the spectral index for each population. In the most recent version (Paper~II) we have included a spectral index distribution in the models. The spectral index for both source classes is normally distributed with mean and standard deviation determined using the maximum likelihood analysis presented in \cite{Venters2007} and data from \cite{Hovatta2014}. In Paper~II, before calculating a flux density, each source is also assigned a random value for the spectral index from the distribution of the corresponding source class.

A major result of Paper~I and Paper~II, which is directly releveant to the work we present in this paper is that the timescale modulation factor (Eq. \ref{tscale_mod}) follows an exponential distribution,
\begin{equation}
P(m)=C({\lambda}e^{-\lambda{m}}),
\label{expo_eq}
\end{equation}
where $C$ is a normalization constant to account for the truncated range and $\lambda$ is the inverse mean of the distribution, with the mean depending on the source class and the flux limit of the sample. Using the upgraded models from Paper~II (which we use throughout this work), the timescale modulation factor has a similar mean (0.381 for the BL Lacs and 0.318 for the FSRQs, for 1.5 Jy flux-density limited samples) for both classes.

\section{Intrinsic Probability Density function of timescales}\label{intr_prob_den_func}

Our aim in this work is to extract information about the intrinsic (rest-frame) probability density function (pdf) of timescales {\em for any class of events characterized by a single timescale per blazar}. Examples of such classes of events include: shortest flare rise time in a blazar; the duration of the largest (in amplitude) flare ever observed in each source; the duration of the longest (in angle) rotation of the polarization angle in each blazar; and the average time delays between flares in two different frequencies in each blazar.

In order to achieve our goals, we first treat the inverse problem: given a rest-frame pdf of timescales in the blazar population for events of a particular class, what would the {\em observed} timescale pdf be, after applying the modulation induced by relativistic effects? In the following, we will use $t_i$ to denote intrinsic (source rest-frame) timescale, and $t_o$ to denote observed (observer-frame) timescales. 

Assuming that the physical processes in the rest frame have no knowledge of redshift or Doppler factor, the intrinsic timescales of any event are independent of the timescale modulation factor $m=(1+z)/D$. As a result,
\begin{equation}
P(t_i,m)=P(t_i)P(m),
\label{intrinsic_pfd}
\end{equation}
where $P(t_i,m)$ is the probability of the intrinsic timescale, $t_i$, to be modified by $m$; $P(t_i)$ is the intrinsic timescale probability density; and $P(m)$ is the timescale modulation factor probability density (Eq. \ref{expo_eq}).
The observed timescales are connected to the intrinsic through $t_o=mt_i$. Thus Eq. \ref{intrinsic_pfd} can be transformed as:
\begin{equation}
P(t_o,m)=P(t_i,m)|J(t_i,m)|,
\label{trans}
\end{equation}
where $P(t_o,m)$ is the probability of observing $t_o$ given $m$, and $J(t_i,m)$ is the Jacobian of the transformation. From the relation between observed and intrinsic timescales, the Jacobian will be: 
\begin{equation}
J(t_i,m)=
\begin{bmatrix}
& \frac{1}{m} & 0\\
& & & \\
& 0 & 1\\
\end{bmatrix}
\end{equation} 
Thus the (absolute value of the) determinant of the Jacobian is equal to $|J(t_i,m)|=1/m$, and Eq.\ref{trans} becomes $P(t_o,m)=P(t_i)P(m)/m$. The probability density function of the observed timescales is thus:
\begin{equation}
 P(t_o)=\int^{m_{max}}_{m_{min}}P(t_o,m)dm=\int^{m_{max}}_{m_{min}}P(t_i)C(\lambda e^{-\lambda m})\frac{1}{m}dm,
\label{pdf} 
 \end{equation} 
where $m_{min}$, $m_{max}$ are the minimum and maximum timescale modulation factors.

Since we know the functional form of $P(m)$ from the results of Papers~I and II, we can assume a family of distributions for $P(t_i)$ and use Eq. \ref{pdf} to obtain the resulting $P(t_o)$. The parameters of $P(t_i)$ can then be recovered by requiring that the resulting $P(t_o)$ fits the observed data best. In this work we provide in Appendix \ref{fam_distr} some examples of this process for the most commonly used and/or encountered distributions in astrophysics in general.

\section{Method Benchmarking}\label{method_app}

In order to benchmark the accuracy of our method under realistic observing conditions, we generate a simulated survey and explore the accuracy of the method with respect to the cadence of observations and sample size.

\subsection{Simulated sample}

For our benchmarking experiments, we have chosen to use the simple case where the class of events under investigation have the same intrinsic characteristic timescale in the rest-frame of all blazars. Then the rest-frame distribution can be described by a $\delta$-function and the observer-frame distribution in the ideal case of perfect sampling will be an appropriately normalized exponential distribution (Eq. \ref{prob_delta}). We constructed our simulated samples by drawing a random value for the timescale modulation factor from an exponential distribution (Eq. \ref{expo_eq}), and multiplying it with the same characteristic intrinsic  timescale (set to $t_i=$100 days to match our real-data application, see \S \ref{robopol_int_time}).  

The minimum modulation factor we accept is $10^{-2}$ following the results of Papers~I and II. The mean of the timescale modulation factor distribution depends on the object class and the flux limit of the sample. For this particular experiment, we choose to use the model for the FSRQ population with an 1.5 Jy flux-density limit ($\lambda^{-1}$ was set to 0.318). Since the mean of the BL Lac population for the 1.5 Jy flux limit is very similar with that of the FSRQs, we expect the conclusions derived for the latter to extend to the former. However, this is not necessarily true for different flux limits. We discuss the possible effects of a different flux limit to the results of our benckmarking in \S \ref{sum & concl}.

\subsection{Effects of sample size and cadence}\label{benchmarking}
\begin{figure}
\resizebox{\hsize}{!}{\includegraphics[scale=1]{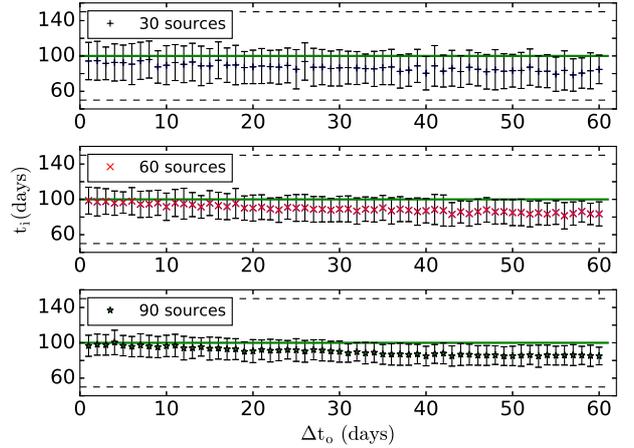} }
 \caption{Estimated best-fit $t_i$ versus interval between observations $\Delta t_o$ (both in days) for samples without pileup. The error bars represent the uncertainty of the best-fit $t_i$. The green line at 100 days shows the value of the ``true'' intrinsic timescale, while the black dashed lines at 50 and 150 days show the limits of the parameter space scanned.}
 \label{plt_time_vs_cad}
 \end{figure} 


\begin{figure}
\resizebox{\hsize}{!}{\includegraphics[scale=1]{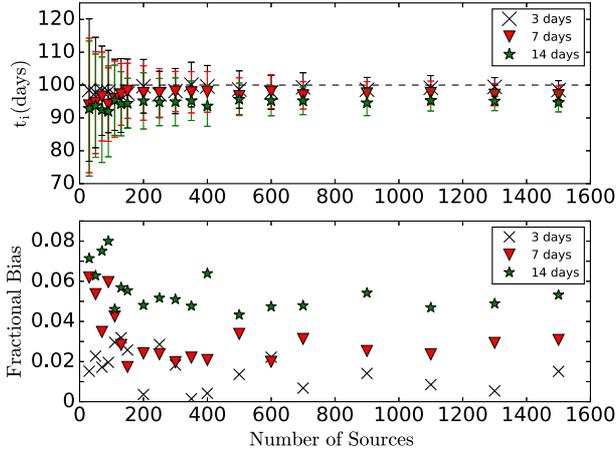} }
 \caption{{\bf Upper panel (a)}: Estimated best-fit $t_i$ versus sample size ranging from 30 to 1500 sources. The error bars represent the uncertainty of the best-fit $t_i$. {\bf Lower panel (b)}: Fractional bias between the estimated best-fit and the ``true'' $t_i$ ($|t_{i,fit}-t_i|/t_i$)  versus sample size. Symbols as above.}
 \label{plt_frac_diff+bias}
 \end{figure}

The finite cadence of any monitoring program can affect measured timescales in three ways. First, it can introduce an artificial cut-off at the lower end of the observed timescale distribution: timescales shorter than $\Delta t_o$ (the time interval between observations) will not be measured depending on $m$. Second, it can also introduce an uncertainty in measured timescales. The measured timescale will lie between $t_{observed}$ and $t_{observed}+2\Delta t_o$. This is because in the worst case scenario, both the beginning/end of an ``event'' can happen exactly after/before the first/last datapoints. For this reason, once we generate our simulated observer-frame sample, we create an ``observed'' sample by drawing a random value from a uniform distribution with range [$t_o-2\Delta t_o$,$t_o$], where $t_o$ is the actual value of the observer-frame timescale.

Third, it may introduce a ``pileup'' of events close to the cadence limit depending on how events faster than $\Delta t_o$ are treated. One possibility is that every timescale smaller than $\Delta t_o$ is rejected as not observed. In this case there is no pileup. Another possibility is that timescales smaller than $\Delta t_o$ are set as observed, with value equal to that of $\Delta t_o$. Physically, the latter case corresponds to observations of events of various timescales (e.g flares in a particular frequency), of which we are interested in the smallest duration event occurring in a particular source. In this scenario faster-than-observed events may occur in a source, but the ``fastest'' one we record is set by the survey cadence, leading to a pileup resulting in a systematic offset in our statistical analysis. For the remainder of this section we will only consider the first possibility (no pileup). For a discussion of the possible effects of pileup, see Appendix \ref{system_effects}.

We perform our benchmarking experiments as follows:

\begin{itemize}

\item{We generate an ``observed'' sample using the procedure described above for a given $\Delta t_o$.}

\item{We construct the cumulative distribution function of Eq. (\ref{prob_delta}) for different assumed values of $t_i$ and compare it with the simulated ``observed'' sample with the use of the Kolmogorov Smirmov test (K-S test). The K-S test provides a straightforward way of rejecting distributions that are inconsistent with our simulated dataset. We accept that for a probability value larger than $p=0.05$ we can not reject the null hypothesis that the two samples are drawn from the same distribution. Throughout this work, we refer to this p-value as the ``probability of consistency''.

In order for our distribution to be normalized in the observer's frame, we need to calculate the normalization constant C (see Appendix A) which depends on $t_{o,min}$ and $t_{o,max}$. In practice, $t_{o,min}$ is established by $\Delta t_o$ and $t_{o,max}$ by the length of the observing season. For our benchmarking experiments, we simulated an observing experiment with an observing season set to ten times the longest observed timescale in the simulated sample. The value of $t_i$ which yielded the highest p-value of the K-S test is the value we considered the best-fit parameter value.}

\item{We repeat the above process 100 times in order to evaluate the statistical spread of the results, for a given sample size and $\Delta t_o$. We quote the mean of these best-fit parameters as the most likely estimate of the rest-frame timescale and the standard deviation of the estimates in different iterations as the uncertainty of the final estimate.}
\end{itemize}

Due to relativistic compression, the shortest rest-frame timescale can either be equal or longer than the shortest one observed. The longest rest-frame timescale can be up to the maximum observed divided by the smallest modulation factor ($\sim 10^{-2}$). Thus in the application of the method, the expected range of the rest-frame timescale estimate should be at least from the shortest to 100 times the longest observed timescale. 

For this particular exercise, since we have a priori knowledge of the $t_i$, the range is set from 50 to 150 days. If the best-fit $t_i$ lies outside of that range, then the estimate will take the value of the closest boundary value. Any best-fit $t_i$ that is either $\leq 50$ or $\geq 150$, would correspond to an at least $50\%$ error in the estimate.

We examined three sample sizes (30, 60, 90 sources) and $\Delta t_o$ starting from 1 day, up to 60 days with a step of 1 day. We draw sources until we reach the desired sample size in the observer's frame ignoring cases where $t_o$ is smaller than $\Delta t_o$.

Figure \ref{plt_time_vs_cad} shows the  estimated best-fit  $t_i$ versus $\Delta t_o$. The error bars are the spread of the estimated rest-frame timescale.  As expected, increasing the number of sources in our samples results in a more narrow spread of estimates. There is a negative bias (systematic shift towards lower values of the estimated intrinsic characteristic timescale) with the increase of the number of days between observations. In many cases for $\Delta t_o\geq 30$ days, the best-fit estimate is more than $1\sigma$ away from the true value. This leads us to the conclusion that for a $\Delta t_o$ of more than 30 days ($\sim$1/3 the actual timescale we are trying to measure), the method loses its ability to accurately estimate the best-fit parameter of the intrinsic probability density function.

In addition, we explore the effect of the size of the sample at a fixed interval between observations. We follow the procedure described above with $\Delta t_o$ fixed to a predefined value, and a varying number of sources in the sample. We chose three values for $\Delta t_o$; 3, 7, and 14 days, all below the 30 days limit determined above (Fig. \ref{plt_time_vs_cad}). The different samples have a size starting from 30 sources up to 150 with a 20 source step, from 150 up to 400 with a 50 source step, from 400 up to 700 with a 100 source step, and from 700 up to 1500 with a 200 source step.

Figure \ref{plt_frac_diff+bias}a shows the estimated $t_i$ versus the sample size with the error bars being the uncertainty of the method, while figure \ref{plt_frac_diff+bias}b shows the fractional bias of the estimated best-fit and the ``true'' $t_i$ ($|t_{i,fit}-t_i|/t_i$) versus sample size. It is obvious that the larger the sample the smaller the spread (Fig \ref{plt_frac_diff+bias}a),  and the smaller the interval between observations, the smaller the bias (Fig. \ref{plt_frac_diff+bias}b). There is no significant change in the fractional bias for a given $\Delta t_o$ regardless of sample size. The method has an accuracy of $\leq 8\%$ in the estimates (Fig. \ref{plt_frac_diff+bias}b) as long as $\Delta t_o \leq 14\%$ of the intrinsic timescale value. For a large number of sources ($\geq 200$) and small $\Delta t_o$ (3 days, i.e 3\% of the timescale we are trying to determine) the bias is smaller than $\sim 1\%$.

\subsection{Testing for the type of distribution}\label{test_distr}
\begin{figure}
\resizebox{\hsize}{!}{\includegraphics[scale=1]{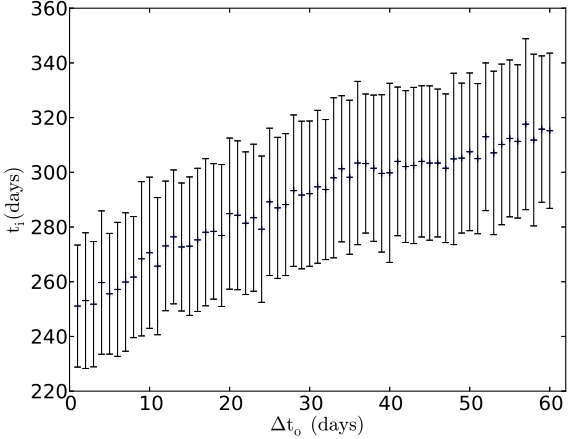}}
 \caption{Estimated $\delta$-function best-fit parameter ($t_i$) for an intrinsic uniform distribution versus $\Delta t_o$. The error bars represent the uncertainty of the best-fit $t_i$.}
 \label{plt_delta_uni}
 \end{figure}  
The best-fit intrinsic distribution family is not necessarily unique. Different families of distributions of intrinsic timescales can produce similar results in the observer's frame. For example, an intrinsic sharp distribution would appear consistent with a $\delta$-function as well as with a narrow normal distribution. Physically this is not necessarily problematic, since all the well-fitting distributions should describe approximately the same physical reality (in our example, a ``preferred'' timescale in the rest-frame). A more worrisome situation would be if, because of uncertainties or systematic effects,  a single dataset appeared consistent with families of distributions describing a very different physical reality. To demonstrate how this effect can cause confusion, we tested whether a finite-width intrinsic distribution can appear consistent with a $\delta$-function. To this end, we performed the following experiment.

We created simulated samples of $t_o$ using an intrinsic uniform distribution from $t_i=$50 to $t_i=$500 days. The sample size was fixed to 200 sources, a size large enough to ensure accurate estimates (as seen in Fig. \ref{plt_frac_diff+bias}b). Using the procedure described in \S \ref{method_app} we fitted an intrinsic $\delta$-function for intervals between observations ($\Delta t_o$) from 1 to 60 days. We assumed no a priori knowledge of the intrinsic distribution or the range of the parameter space. Figure \ref{plt_delta_uni} shows the $\delta$-function best-fit parameter $t_i$ versus $\Delta t_o$. The best-fit estimates of the characteristic timescale fall approximately at the median value of the intrinsic distribution with a systematic shift towards higher values for increasing time intervals between observations. The K-S test yields a probability of consistency that ranges from $45\%$ to $\leq 65\%$. In this case we could naively assume that all blazars have the same characteristic timescale in the jet rest frame. 

However, there is a simple test for this particular case that can clarify the nature of the intrinsic distribution. If indeed the intrinsic distribution of timescales is a $\delta$-function, or at least a sharp distribution, then by fitting a normal distribution, we should in principle find that the best-fit distribution is very sharp. For simplicity we tested only two intervals, 7 and 30 days. The 7 day interval was chosen because it is the limit at which the method becomes very accurate ($\leq$ 4\% error) and the 30 day interval because it is the limit after which results are unreliable (see Fig. \ref{plt_time_vs_cad}).

For the 7 day interval we find that the best-fit intrinsic normal distribution would have mean $\mu=253.0$ and standard deviation $\sigma=168.0$ with a 97\% probability of consistency between samples. For the 30 day interval the best-fit normal distribution has $\mu=168.8$ and $\sigma=239.0$ with a 76\% probability of consistency. In both cases the resulting best-fit normal distribution is far from consistent with a $\delta$-function. The fact that they are significantly wider and with more than 30\% higher probability of consistency is suggestive that the ``true'' intrinsic distribution is much wider than a $\delta$-function. Such simple tests can provide invaluable insights on the ``true'' shape of the intrinsic distribution in the application of  the method.

\subsection{K-S test versus conventional fitting methods}\label{test_MLE}
\begin{figure}
\resizebox{\hsize}{!}{\includegraphics[scale=1]{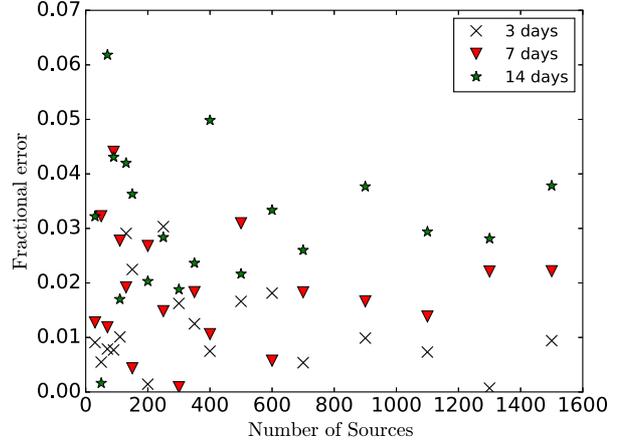}}
 \caption{Fractional error between the maximum likelihood and K-S test fitting methods versus sample size  for the 3 day, the 7 day, and the 14 day time interval between observations ($\Delta t_o$).}
 \label{plt_MLE_diff}
 \end{figure}
The K-S test provides a convenient way to automatically reject families of rest-frame distributions that are a poor fit to the data, especially in cases when there is no analytical solution. However, the K-S statistic (or the associated p-value) are not formally appropriate for fitting. To examine how much error we introduce by optimizing our parameters through K-S statistic minimization, we perform the same analysis in section \ref{benchmarking} for fixed time intervals between observations, but instead of using the K-S test fitting method, we use a maximum likelihood estimation (MLE) of parameters. Figure \ref{plt_MLE_diff} shows the fractional error between the K-S statistic and maximum likelihood fitting methods. Even for the longest time interval (14 days) between observations, the fractional error is $\lesssim 6\%$. In cases with small $\Delta t_o$ and a large number of sources, the fractional error is only $\sim 1\%$. We thus conclude that the K-S test method used throughout this work can adequately mimic a formally appropriate fitting method and provide robust results without the computational cost and complexity required by other methods. For a discussion of K-S versus MLE fitting in the case of population models themselves, see Paper~I.

\section{Application to blazar polarization angle swings}\label{robopol_int_time}
\begin{figure}
\resizebox{\hsize}{!}{\includegraphics[scale=1]{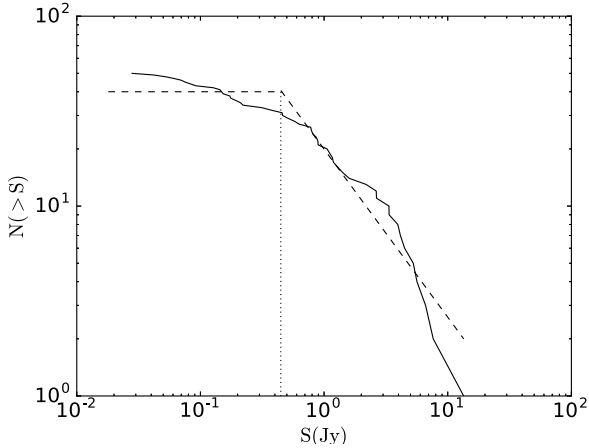} }
 \caption{Source count distribution (number of sources N with flux density larger than S) for the RoboPol main sample sources. The black dotted line marks the assumed radio flux-density limit.}
 \label{plt_robopol_flux-limit}
 \end{figure}

\begin{figure}
\resizebox{\hsize}{!}{\includegraphics[scale=1]{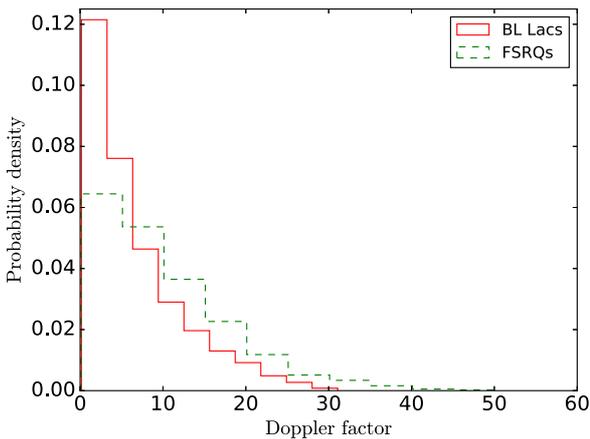}}
 \caption{Distribution of the simulated Doppler factors for the RoboPol flux-density limited sample. Solid red is for the BL Lacs, and dashed green is for the FSRQs.}
 \label{plt_doppler_robopol}
 \end{figure} 
\begin{figure}
\resizebox{\hsize}{!}{\includegraphics[scale=1]{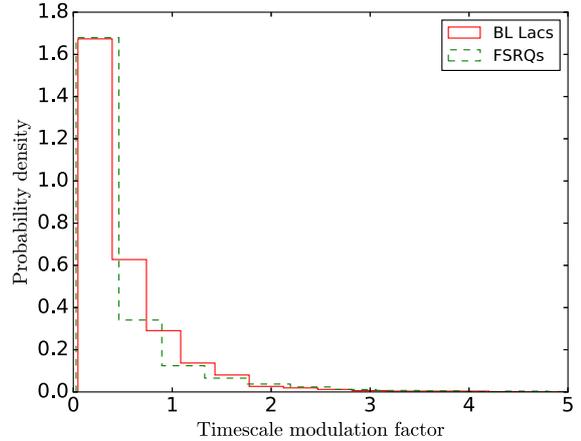} }
 \caption{Distribution of the simulated timescale modulation factor for the RoboPol flux-density limited sample. Solid red is for the BL Lacs, and dashed green is for the FSRQs.}
 \label{plt_mod_robopol}
 \end{figure} 
 
\begin{figure}
\resizebox{\hsize}{!}{\includegraphics[scale=1]{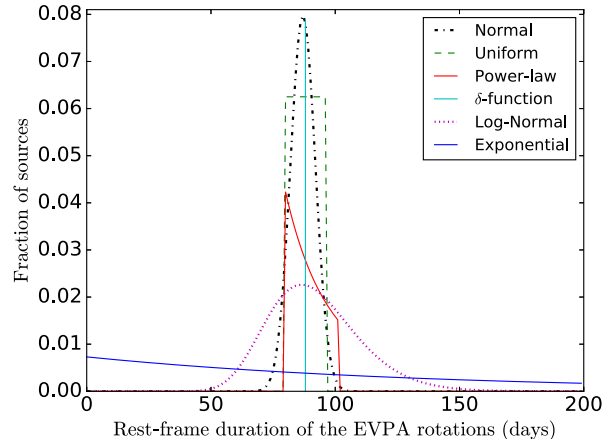}}
 \caption{Best-fit distribution, for various distribution families,  of the intrinsic timescales of the longest EVPA rotations in a radio flux-density limited subsample of RoboPol blazars.}
 \label{plt_intrinsic_time_distr}
 \end{figure}

Having benchmarked our method, we present the first real-data application to timescales associated with the rotation of the polarization plane (Electric Vector Position Angle (EVPA) rotation), in optical, seen in blazars. The physical mechanism of these rotations is, to this day, unknown, with models ranging from random walk processes \citep{Moore1982,Marscher2014,Kiehlmann2016} to shocks propagating in the helical magnetic field of the jet \citep{Marscher2008,Marscher2010} and bends in the jet \citep{Abdo2010-III}. Although some rotations have been associated with $\gamma-ray$ flares and ejections of radio components \citep{Marscher2008,Marscher2010} this is not always the case \citep{Blinov2015}.

In order to uncover the underline intrinsic timescale distribution of these rotations, we use data from the RoboPol survey \citep{King2014}. The RoboPol survey uses a $\gamma$-ray flux-limited sample, specifically designed for rigorous statistical studies \citep{Pavlidou2014,Angelakis2016}. Since our method works for radio flux-density limited samples, the first step to applying our method is to identify an appropriate subset of the RoboPol main sample that is an unbiased (randomly drawn) subsample of a radio flux-density limited sample. All the RoboPol sources have been monitored by the Owens Valley Radio Observatory (OVRO) blazar program \citep{Richards2011}. We use the maximum likelihood mean flux-densities \citep{Richards2014} to construct the source count distribution for the RoboPol main sample sources (N(>S) as a function of S, Fig \ref{plt_robopol_flux-limit}) and we approximate it with a horizontal part at low flux densities S and a declining power-law at higher S. We take sources brighter than 0.446 Jy (the transition point between the two approximations, see Fig. \ref{plt_robopol_flux-limit}) to be the desired subsample, which we consider to be approximately unbiased compared to a parent flux-density-limited sample with a limit of 0.446 Jy. We have verified that our results are not sensitive to the exact location of the flux-density limit (i.e the inclusion or the omission of one more source). The number of sources in our radio flux-density limited sample is 31.

Figure \ref{plt_doppler_robopol} shows the simulated distribution of Doppler factors and Fig. \ref{plt_mod_robopol} the simulated distribution of the timescale modulation factor for a 0.446 Jy radio flux-density limit, as derived from the population models in Paper II. Due to the low flux-density limit, we expect that the Doppler factor values will be, on average, smaller than the ones derived from VLBI or variability studies of brighter sources \citep{Hovatta2009,Liodakis2016} and the ones in the higher flux-density--limit simulated samples we investigated in Papers~I and~II. The mean is $\sim 6$ for the BL Lacs and $\sim 10$ for the FSRQs. The mean of the timescale modulation factor is 0.475 for the FSRQs, and 0.49 for the BL Lacs. Since the two values are very similar, we adopted their mean (0.4825) as the common value. This way we avoid splitting the sample and reducing our statistics. We have verified that using either time scale modulation factor mean (FSRQ or BL Lac population) for the value of the whole sample does not result in any significant change on our results or our conclusions.

\subsection{Observations}

The data were collected during the 2013, 2014 and 2015 observing seasons of the RoboPol survey. The definition of the EVPA swings we adopt is similar to the definition of an EVPA rotation from \cite{Blinov2015}. We define an EVPA rotation as any continuous change of EVPA with total amplitude $\Delta \theta_{total}> 90^o$  comprised of at least four consecutive observations with statistically significant swings ($\Delta\theta>\sqrt{\sigma^2_{\theta_{i+1}}+\sigma^2_{\theta_{i}}}$, where $\sigma_\theta$ is the uncertainty in the EVPA) between them. The start and end points of a rotation are defined by a factor of 5 change of the slope of the EVPA time series, $\Delta \theta / \Delta t$ or a change of its sign. Contrary to \cite{Blinov2015} we do not apply any limits in the number of points or the length (rotation angle $\Delta \theta_{total}$) of an event. Following the definition in \cite{Blinov2015} we find 29 rotations in 14 sources, whereas with the definition adopted in this work we find 570 rotations in 31 sources making a statistical approach feasible.

Our method is applicable when each blazar is characterized by a single observed timescale. In the case of EVPA rotations, for each blazar we select the rotation (as defined above) with the longest duration (longest rotation timescale $\Delta T_{max}$).  The shortest time interval between observations, $\Delta t_o$,  is $\sim$1 day with an average of $\sim$9 days, while the events (see Fig. \ref{plt_robopol_normal_uniform}) last from 3 to 108 days.

\subsection{Results}
\begin{figure}
\resizebox{\hsize}{!}{\includegraphics[scale=1]{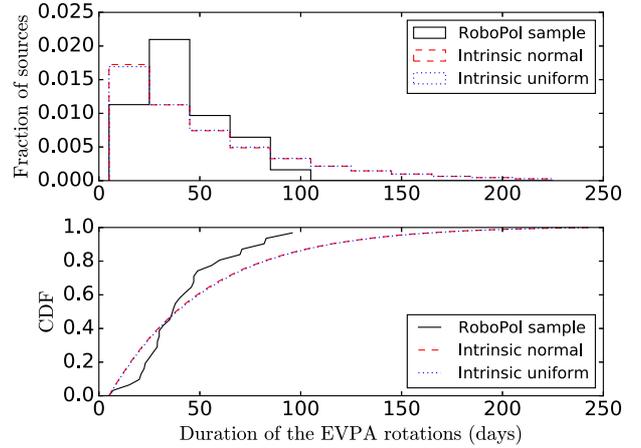} }
 \caption{{\bf Upper panel:} Distribution of $\Delta T_{max}$ in our sample. {\bf Lower panel:} Cumulative distribution function of the $\Delta T_{max}$ of the EVPA rotations.}
 \label{plt_robopol_normal_uniform}
 \end{figure} 

Following the procedure described in \S \ref{method_app}, we apply our method to the radio flux-density limited RoboPol subsample testing the six families of distributions described in Appendix \ref{fam_distr}. For the intrinsic timescales, we explored a parameter space from 1 day to 10$\times$ the maximum observing season length (244 days) of the RoboPol survey. In the case of the power-law index the range of the parameter space was set to [-7,7]. The best fit parameter values for each family of distributions and the K-S test p-value for each are shown in Table \ref{tab:Parameter values}. The uncertainties in the parameters given in Table \ref{tab:Parameter values} indicate the range in which for those parameters the K-S test yielded a $>5\%$ probability of consistency between observed and simulated samples.
\begin{table}
\setlength{\tabcolsep}{12pt}
\centering
  \caption{Best-fit parameters and probability of consistency (K-S test) values for each family of distributions. The uncertainties indicate the range of the parameter that can produce an acceptable fit (i.e K-S test probability $>5\%$). In the case of the log-normal distribution, $t_i$ and $\sigma_{t_i}$ are related to $\mu$ and $\sigma_{sc}$ (see \ref{fam_distr}) through $t_i=exp(\mu+\sigma_{sc}^2/2)$ and $\sigma_{t_i}=\sqrt{ exp(\sigma_{sc}^2-1)exp(2\mu+\sigma_{sc}^2)}$}
  \label{tab:Parameter values}
\begin{tabular}{@{}ccccc@{}}
 \hline
    Distribution  & $t_i$ &  $\sigma_{t_i}$ &  & p-value  \\
   & (days) &  (days) &{} & (\%)  \\
  \hline
    Normal & $87_{-15}^{+26}$   & $5_{-4}^{+35}$ & {} & 31.5  \\[2pt]
   
    $\delta$-function &  $88_{-19}^{+16}$  & - &{} & 29.7 \\[2pt]
     
    Log-Normal & $92_{-18}^{+89}$  & $18_{-17}^{+220}$ & {}& 25.5 \\[2pt]
    Exponential  & $137_{-15}^{+28}$  & - & {} & 11.8 \\
 \hline
    Distribution  & $t_{i,min}$ &  $t_{i,max}$ & slope  & p-value  \\
   & (days) & (days) & & (\%)  \\
  \hline   
    Uniform & $80_{-10}^{+14}$ & $96_{-16}^{+33}$ & - & 31 \\[2pt]  
    Power-law & $80_{-20}^{+20}$ & $101_{-21}^{+399}$ & $-4.4_{-1.1}^{+0.8}$ & 30 \\
\hline
\end{tabular}
\end{table}

All families of distributions, with the exception of the exponential, converge to approximately the same range of intrinsic timescales (Fig. \ref{plt_intrinsic_time_distr}) with very similar K-S test probabilities. The probability of consistency is highest for the normal and uniform distributions (31.5\% and 31\% respectively). For these two intrinsic (rest-frame) distributions, it is clear even by inspection that the resulting observer-frame distributions are almost indistinguishable (Fig \ref{plt_robopol_normal_uniform}). We observe a lack of very short and very long duration events. Both however can be attributed to observational constrains. For example, the former could be due to the cadence being too low to observe shorter smooth variations. Indeed, there are blazars for which even the longest observed smooth variation does not last much more than the time interval between observations. The current (season 2016) RoboPol observing strategy (i.e monitoring of the most variable sources with nightly observations, $\Delta t_o=1$)  will be able to resolve faster (shorter duration) events.

The latter could be due to the length of the observing seasons ($\sim 240$ days) or due to the source not being available for observations throughout the observing season. In Appendix \ref{season_len}, we test whether observational constrains could significantly alter our results by shortening long-duration events. We find that it is highly unlikely that the observed event timescale distribution is significantly wider than the observed, thus our results are independent of such constrains.

For the best-fit intrinsic $t_i$ (87 days) we estimate, based on our benchmarking of \S \ref{benchmarking}, that the bias is $\leq$8\% since the average $\Delta t_o$ is only $\sim$10\% of the rest-frame $t_i$. From Fig. \ref{plt_frac_diff+bias}b we can conclude that the results of our analysis are robust independent of our sample size since the average $\Delta t_o$ is $<14\%$ of the $t_i$. Given that we are looking for the longest, rather than the shortest, event timescale in each blazar, a systematic pileup of timescales close to the average distance between observations (see Appendix \ref{system_effects}) is not expected in this particular application. It should be noted that although the narrowly peaked intrinsic distributions have a systematically higher probability of consistency than the much wider exponential distribution, the exponential remains an acceptable fit, and our sample size is not large enough to conclusively reject a significantly wider distribution of rest-frame timescales  (it is clear from Table \ref{tab:Parameter values} that the lognormal and power-law distributions can also yield acceptable fits for a wider distribution).  However, the exponential distribution is a monoparametric family, and none of the multi-parametric families prefers a shape similar to the best-fit exponential, indicating a preference of the data for the narrower distributions, according to the test we discussed in \S \ref{test_distr}.

To check whether the ``true'' intrinsic distribution is an exponential, and our results arose from random sampling, we performed the following test. We created $10^4$ simulated datasets from an intrinsic exponential distribution with the same mean as in Table \ref{tab:Parameter values} in a simulated observing survey with the same characteristics ($\Delta t_o$, observing length etc.) as RoboPol. We then used the distributions discussed above (including an exponential) with the same parameters from Table \ref{tab:Parameter values} to fit the simulated samples and calculate the K-S test statistic. Out of the $10^4$ simulated samples, we calculated the percentage of trials that achieved a K-S test p-value equal or higher than the one reported in Table \ref{tab:Parameter values} for each family of intrinsic distributions (other than the exponential). The probability values  range between $<10^{-4}$ and 25\%.  However, there was no case where at the same time an exponential distribution would yield a $<11.8\%$ probability while {\em all the other families} yield a $>25.5\%$. For this reason, we formally reject the exponential distribution as a acceptable model.

\subsection{Possible interpretations}

Our findings suggest that the intrinsic timescales of the longest EVPA events in the radio flux-density limited subsample of RoboPol blazars that we have examined are confined to a relatively small range (spread of intrinsic timescales $\sim 30$ days).
 It is possible that a subgroup in our sample has a small range of intrinsic timescales and is dominating over the rest of the sample.

We have tested for this scenario in the following way. We simulated a physical situation where our sample was an admixture of sources with a narrow distribution of timescales (distributed according to the best-fit uniform distribution describing the RoboPol sample) and sources with a wide distribution of timescales (uniform [50,500] days). We created a simulated sample following the procedure described in \S \ref{method_app} and proceeded in fitting an intrinsic uniform distribution. We repeated the same procedure 31 times, and each new simulated sample had a larger (by one) number of sources that were drawn from the wider distribution. Even with a small number of sources from a wide uniform distribution (<5 sources) the method is unable to produce a distribution with such a small range. We conclude that if there were two underlining distributions of significantly different widths, the wider of the two would dominate even if it only contributed a small number of sources.

If we assume that the EVPA rotations are produced by a shock propagating downstream in a jet, we should observe a rotation when its tracing the spirals of the helix of the jet \citep{Marscher2008,Marscher2010}. In this case our results will imply that the maximum length of the jet traveled during a rotation is similar for all blazars and is roughly between 75 and 100 light days. In this case, the sources would have a narrow distribution of intrinsic timescales but due to different Lorentz factors and viewing angles the observed distribution is much wider. This would suggest that the intrinsic size of the jets is very similar for all blazars. However, given the range of jet sizes seen in radio galaxies (i.e the unbeamed parent distribution of blazars) this is highly unlikely.

If the EVPA rotations are caused by a random walk processes, the range of intrinsic timescales found in this work would represent a typical timescale of the smooth variation of turbulent plasma cells. To test if random walk processes can produce such a small range of timescales we perform the following experiment. We use the random walk model described in \cite{Blinov2015} to create a simulated EVPA curve comparable to the EVPA curve of each source. We create a simulated distribution of the longest EVPA rotations and we repeat the process to recover the intrinsic timescale distribution as described above. The simulated sample created via the random walk process has a 56\% probability of consistency with the observed according to the K-S test. We found that the best-fit distribution is a normal distribution with mean 69 days and standard deviation 5 days with a 82\% probability of consistency followed by a uniform with $t_{i,min}=66$ and $t_{i,max}=73$ and 81.5\% probability of consistency. The results are very similar, showing that the random walk process can indeed reproduce such a small range and cannot be ruled out as a potential mechanism. It is interesting that the probability of consistency is more than twice as much as the one of the observed data. This could be coincidental, due to the random processes responsible for the simulated distribution, or could indicate the existence of two EVPA rotation mechanisms. It is argued in \cite{Blinov2015} and \cite{Kiehlmann2016} that not all rotations can be explained by a random walk or a deterministic event. Therefore it is not unlikely that rotations created by random walks dominate our observed sample, yet there are still deterministic events present creating the large difference between K-S test p-values of random walk simulated and observed samples.
 
The assumption implied in treating the RoboPol data is that the beaming between the optical and radio regions is the same. If this is not the case, then the timescale modulation factor distribution would change, altering our results. The fact that the intrinsic distribution is narrow suggests that the difference in beaming is likely to be systematic, with all sources having either higher or lower Doppler factors. Otherwise the induced spread would create the appearence of a much wider distribution. If this is indeed the case, the width of the best-fit distributions will not change, but the distribution will be shifted to higher or lower values.  Finally, we caution that, although disfavoured by the data, distributions allowing for a much wider range of rest-frame timescales are not formally rejected by our relatively small source sample (Table \ref{tab:Parameter values}).

\section{Summary and Conclusions}\label{sum & concl}

In Papers~I and II, we found that the distributions of observer-frame timescales of the blazar populations are modulated by an exponential distribution compared to their rest-frame counterparts. This exponential has a mean depending on source class and the flux-density limit of the sample.

In this work, assuming that the intrinsic timescales of blazars are independent of redshift and Doppler boosting, we have developed a novel method of uncovering the underlying rest-frame probability density function of timescales of blazar jets using the observer-frame probability density.

We caution that the independence of rest-frame timescales from redshift and Doppler boosting is not obvious. There may be a dependence on either redshift or Doppler boosting, or both. However such a dependence can best be addressed in a model-dependent fashion (i.e. by testing an explicit model proposing such a relationship). Our formalism can be extended in a straightforward fashion to test such models.

In addition we have benchmarked  our  method in a realistic observing scenario using the timescale modulation factor distribution for the FSRQ population with a flux-density limit of 1.5 Jy and we assessed the impact of various systematic effects.

We found that when the interval $\Delta t_o$ between observations is longer than 30\% of the timescale we try to measure, there is a systematic negative bias in the estimates that can lead to best-fit parameters deviating more than 1$\sigma$ from the characteristic timescale. Low cadence of observations can thus prevent our method from producing robust results.

Exploring the method's sensitivity to sample size we found that all estimates for sample sizes $\geq 30$ sources have an accuracy of $\leq 8\%$ if $\Delta t_o$ is sufficiently low. Monitoring programs with small $\Delta t_o$ and large number of sources such as the Owens Valley Radio Observatory (OVRO)\footnote{http://www.astro.caltech.edu/ovroblazars} blazar monitoring program \citep{Richards2011} would make an ideal candidate for the application of our method.

In practice, observational constraints of specific experiments may induce additional effects other than the ones discussed in our general benchmarking. For this reason, we encourage the users of our method to conduct additional program-specific simulations that can uncover such effects, as we have also done in Appendix \ref{season_len} to address the specific effect of interrupted seasonal observations on EVPA rotation lengths.

The family of distributions that best fits a particular observer-frame dataset does not need to be unique. Different families can provide similarly fitting results. A test that can help determine whether the deduced rest-frame distribution is an acceptable description of the physical reality of the source population is to compare the best fits from different families of distributions. All distributions with the same number of parameters should converge to a similar answer (for example, regarding distribution width) for the obtained fit to be considered reliable. Distributions with more parameters should converge to the simpler distribution shapes, if the simpler distributions are acceptable descriptions of reality, otherwise the simpler distributions should be rejected.

The results of the analysis presented in this work during the benchmarking of our method are representative for a simulated 1.5 Jy flux-density limited FSRQ sample. For the same flux-density limit the mean of the timescale modulation factor distribution for the BL Lac population is fairly similar. Thus we would not expect any significant differences with the conclusions derived for the FSRQ sample. However, for a sample with a different flux-density limit the overall amount of beaming will be in principle different for different classes. The flux-density limit will affect the regimes of robustness of our method.

Samples with flux-density limit lower than 1.5 Jy will have a larger mean of the timescale modulation factor distribution, whereas a higher flux-density limit  will result in a smaller mean. In the first case (lower flux-density limit) there will be a shift towards higher values of the observed timescales which in turn will push the limit ($\Delta t_o=30$ days) for the method to produce robust results to higher values. The opposite effect will be true for the higher flux-density limit. Shorter observed timescales will bring that limit to smaller $\Delta t_o$ values.

We have applied our method to the maximum duration of optical EVPA rotations observed in each blazar in a radio flux-density limited sample observed by RoboPol. We have found that the best-fit intrinsic distribution is a normal distribution with mean 87 days and standard deviation 5 days with a 31.5\% probability of consistency between observed and simulated samples. An intrinsic uniform distribution with $t_{i,min}=80$ and $t_{i,max}=96$ has a 31\% probability of consistency making it an equally preferred candidate since the two simulated observer-frame distributions are indistiguisable (Fig. \ref{plt_robopol_normal_uniform}). We have examined several interpretations of our results. If a significant fraction of our events are the result of a random-walk process, a result similar to the one of this work would be obtained. However, without better statistics we cannot exclude other interpretations.

In the case of timescale distributions in different wavelengths (e.g optical, $\gamma$-rays), whether the same boosting (Doppler factor) applies is still an open question. Deceleration in the jet from the $\gamma$-ray to the radio emission region \citep{Georganopoulos2003-I,Georganopoulos2003-II,Georganopoulos2004,Giannios2009} may result in the underestimation of the intrinsic timescales. If such a deceleration does not exist or its effect is not significant, timescales derived from the Fermi Gamma-Ray Space observatory all sky survey \citep{Acero2015} will also constitute an ideal dataset for the application of our method due to the high cadence of observations and sample size, although one should keep in mind that a flux-density limited sample in $\gamma$-rays does not translate directly into a flux-density limited sample in radio \citep{Pavlidouff}, which is the basis of our population models.

On the other hand, applications of our method to timescales extracted by radio monitoring of flux-density limited samples do not suffer from either of these potential problems, and thus constitute prime candidates where our method could be applied with maximum confidence, provided that the cadence of the monitoring is sufficiently high.

\section*{Acknowledgments}
The authors would like to thank Talvikki Hovatta, Dimitrios Giannios and Sebastian Kiehlmann for comments that helped improve this work. This research was supported by the ``Aristeia'' Action of the  ``Operational Program Education and Lifelong Learning'' and is co-funded by the European Social Fund (ESF) and Greek National Resources, and by the European Commission Seventh Framework Program (FP7) through grants PCIG10-GA-2011-304001 ``JetPop'' and PIRSES-GA-2012-31578 ``EuroCal''.

\bibliographystyle{mnras}
\bibliography{bibliography} 

\appendix
\section{Observed Probability Density functions}\label{fam_distr}

\subsection{Intrinsic Delta Function Distribution}
If the intrinsic timescale distribution is a Delta function (all events share the same duration in the rest frame in all blazars) in the form of $\delta(t_{i}-\frac{t_o}{m})$, where $t_{i}$ is the characteristic timescale of the $\delta$-function, Eq. \ref{pdf} becomes:
\begin{equation}
P(t_o)=\int^{m_{max}}_{m_{min}}\delta(t_{i}-\frac{t_o}{m})C({\lambda}e^{-\lambda{m}})\frac{1}{m}dm.
\end{equation}
In order to solve the integral we have to transform $\delta(t_{i}-\frac{t_o}{m})$ to $\delta(m-\frac{t_o}{t_{i}})$ taking into account that, in general,
\begin{equation}
\delta(g(x))=\frac{\delta(x-x_0)}{|g'(x)|_{x_0}|}.
\label{delta}
\end{equation}
In our case $g(m)=t_{i}-\frac{t_o}{m}$ and $\delta(m-x_0)=\delta(m-\frac{t_o}{t_{i}})$. From Eq. \ref{delta} we have:
\begin{equation}
\delta(t_{i}-\frac{t_o}{m})=\frac{t_o}{t_i^2}\delta(m-\frac{t_o}{t_i})
\end{equation}
The probability density function of the observed timescales will be:
\begin{eqnarray}
P(t_o)&=&\int^{m_{max}}_{m_{min}}\frac{t_o}{t_i^2}\delta(m-\frac{t_o}{t_i})C({\lambda}e^{-\lambda{m}})\frac{1}{m}dm\nonumber\\
&=&\frac{C\lambda}{t_i}{e}^{-\lambda(t_o/t_i)}.
\end{eqnarray}
The value of C can be calculated as a function of $t_i$ and $t_{o,min}$, $t_{o,max}$ (the bounds of the observed timescales)  by requiring $P(t_o)$ to be normalized:
\begin{equation}
C=\frac{1}{t_i[e^{-\lambda{t_{o,min}}/t_i}-e^{-\lambda{t_{o,max}}/t_i}]}.
\end{equation}
Thus the probability density function will be:
\begin{equation}
P(t_o)=\frac{\lambda{e}^{-\lambda(t_o/t_i)}}{t_i[e^{-\lambda{t_{o,min}}/t_i}-e^{-\lambda{t_{o,max}}/t_i}]}.
\label{prob_delta}
\end{equation} 
Equation \ref{prob_delta} can be fitted to the observed data,  in order to optimize the characteristic timescale $t_i$ of the intrinsic $\delta$-function

\subsection{Intrinsic Uniform Distribution}\label{math_uniform}
If all timescales in blazar jets in the rest-frame are equally probable, we can assume an intrinsic uniform distribution in the form of:
\begin{equation}\label{intri_unif}
P(t_i)=\left\{
\begin{tabular}{lr}
$\frac{1}{t_{i,max}-t_{i,min}}$, & $t_{i,min}\leq t_i\leq t_{i,max}$\\
$0$,& $ t_i\leq t_{i,min}$ or $t_i\geq t_{i,max}$
\end{tabular}\right.
\end{equation}
Using the Heaviside step function defined as:
\begin{equation}\label{heaviside}
 H(x)=\left\{
\begin{tabular}{lr}
$0$, & $x<0$\\
$1$,& $ x>0$
\end{tabular}\right.
\end{equation}
we can re-write Eq. \ref{intri_unif}:
\begin{eqnarray}
P(t_i)&=&\frac{1}{t_{i,max}-t_{i,min}}H(t_i-t_{i,min})H(t_{i,max}-t_i)\nonumber\\
&=&\frac{1}{t_{i,max}-t_{i,min}}H(\frac{t_o}{m}-t_{i,min})H(t_{i,max}-\frac{t_o}{m}).
\label{intri_unif2}
\end{eqnarray}
Then Eq. \ref{pdf} becomes:
\begin{eqnarray}
P(t_o)&=&\frac{C\lambda}{t_{i,max}-t_{i,min}}\int^{m_{max}}_{m_{min}}\frac{1}{m}e^{-\lambda m}\nonumber\\
&\times &H(\frac{t_o}{m}-t_{i,min})H(t_{i,max}-\frac{t_o}{m}) dm.
\label{prob_uni}
\end{eqnarray}
Due to the properties of the Heaviside step function (Eq. \ref{heaviside}) the observed probability density will be non-zero for $\frac{t_o}{m}-t_{i,min}>0\Rightarrow m<\frac{t_o}{t_{i,min}}$ and $t_{i,max}-\frac{t_o}{m}>0\Rightarrow m>\frac{t_o}{t_{i,max}}$. For the bounds of the integral there are four cases, two for the upper and two for the lower bound. For the upper bound either $m_{max}>\frac{t_o}{t_{i,min}}$ and the bound is $\frac{t_o}{t_{i,min}}$, or 
$m_{max}<\frac{t_o}{t_{i,min}}$ and the bound is $m_{max}$. For the lower bound either $m_{min}>\frac{t_o}{t_{i,max}}$ and the lower bound is $m_{min}$, or $m_{min}<\frac{t_o}{t_{i,max}}$ and the lower bound is $\frac{t_o}{t_{i,max}}$:

If $m_{max}<\frac{t_o}{t_{i,min}}$ then:
\begin{equation}
 P(t_o)=\left\{
\begin{tabular}{lr}
$\frac{C\lambda}{t_{i,max}-t_{i,min}}\int^{m_{max}}_{m_{min}}\frac{1}{m}e^{-\lambda m}dm$, & $t_o\leq t_{i,max}m_{min}$\\
 & \\
$\frac{C\lambda}{t_{i,max}-t_{i,min}}\int^{m_{max}}_{t_o/t_{i,max}}\frac{1}{m}e^{-\lambda m} dm$,& $ t_o\geq t_{i,max}m_{min}$
\end{tabular}\right.
\label{prob_uni2}
\end{equation}

If $m_{max}>\frac{t_o}{t_{i,min}}$ then:
\begin{equation}
 P(t_o)=\left\{
\begin{tabular}{lr}
$\frac{C\lambda}{t_{i,max}-t_{i,min}}\int^{t_o/t_{i,min}}_{m_{min}}\frac{1}{m}e^{-\lambda m}dm$, & $t_o\leq t_{i,max}m_{min}$\\
 & \\
$\frac{C\lambda}{t_{i,max}-t_{i,min}}\int^{t_o/t_{i,min}}_{t_o/t_{i,max}}\frac{1}{m}e^{-\lambda m} dm$,& $ t_o\geq t_{i,max}m_{min}$
\end{tabular}\right.
\label{prob_uni3}
\end{equation}

Fitting Eq. \ref{prob_uni2} and \ref{prob_uni3} to the observed data, we can optimize for $t_{i,min}$ and $t_{i,max}$  that enter Eq. \ref{intri_unif2}.

\subsection{Intrinsic Power Law Distribution}

We now assume that the intrinsic timescales in all blazars follow a power law distribution with slope k in the form of:
\begin{equation}
P(t_i)=C_1 t_i^{k}H(t_i-t_{i,min})H(t_{i,max}-t_i).
\label{intri_power}
\end{equation}
$H(t_i-t_{i,min})$ and $H(t_{i,max}-t_i)$ are Heaviside step functions (Eq. \ref{heaviside}) in order to account for the truncated range of the intrinsic timescales. Equation \ref{pdf} becomes:
\begin{equation}
P(t_o)=\int^{m_{max}}_{m_{min}}C_1 (\frac{t_o}{m})^{k}H(\frac{t_o}{m}-t_{i,min})H(t_{i,max}-\frac{t_o}{m})C(\lambda e^{-\lambda m})\frac{1}{m}dm
\end{equation}
Setting $C_2=C_1C$, and following the procedure described in \S \ref{math_uniform}:

If $m_{max}<\frac{t_o}{t_{i,min}}$ then:
\begin{equation}
 P(t_o)=\left\{
\begin{tabular}{lr}
$C_2\lambda t_{o}^k\int^{m_{max}}_{m_{min}}\frac{1}{m^{k+1}}e^{-\lambda m}$, & $t_o\leq t_{i,max}m_{min}$\\
 & \\
$C_2\lambda t_{o}^k\int^{m_{max}}_{t_o/t_{i,max}}\frac{1}{m^{k+1}}e^{-\lambda m} $,& $ t_o\geq t_{i,max}m_{min}$
\end{tabular}\right.
\label{prob_power2}
\end{equation}

If $m_{max}>\frac{t_o}{t_{i,min}}$ then:
\begin{equation}
 P(t_o)=\left\{
\begin{tabular}{lr}
$C_2\lambda t_{o}^k\int^{t_o/t_{i,min}}_{m_{min}}\frac{1}{m^{k+1}}e^{-\lambda m}$, & $t_o\leq t_{i,max}m_{min}$\\
 & \\
$C_2\lambda t_{o}^k\int^{t_o/t_{i,min}}_{t_o/t_{i,max}}\frac{1}{m^{k+1}}e^{-\lambda m} $,& $ t_o\geq t_{i,max}m_{min}$
\end{tabular}\right.
\label{prob_power3}
\end{equation}

We can thus fit a function of the form given in equations \ref{prob_power2} and \ref{prob_power3} to the observed data to obtain the optimal $k$, $t_{i,min}$, and $t_{i,max}$ that enter Eq. \ref{intri_power}.

\subsection{Intrinsic Exponential Distribution}
Assuming that the intrinsic timescales in blazar jets follow an exponential distribution in the form of:
\begin{equation}
P(t_i)=C_3\nu e^{-\nu t_i},
\label{intri_exp}
\end{equation}
where $C_3$ is the normalization constant, and $\nu$ is the inverse mean of the distribution, equation \ref{pdf} becomes:
\begin{equation}
P(t_o)=C C_3\nu\lambda\int^{m_{max}}_{m_{min}}\frac{1}{m}\exp\left[-(\lambda m+\nu t_o/m)\right] dm
\label{prob_expo}
\end{equation}
Fitting Eq. \ref{prob_expo} to the observed data we can optimize $\nu$ that enters Eq. \ref{intri_exp}.

\subsection{Intrinsic Normal and Log-Normal Distributions}

Assuming that the intrinsic timescales in all blazars are normally distributed with mean $\mu$, and standard deviation $\sigma$ in the form of:
\begin{equation}
P(t_i)=\frac{1}{\sigma\sqrt{2\pi}}\exp\left[-\frac{(t_i-\mu)^2}{2\sigma^2}\right],
\label{intri_normal}
\end{equation}
equation \ref{pdf} becomes:
\begin{eqnarray}
P(t_o)&=&C\lambda\int^{m_{max}}_{m_{min}}\frac{1}{\sigma\sqrt{2\pi}}\exp\left[-\frac{(\frac{t_o}{m}-\mu)^2}{2\sigma^2}-\lambda m\right]\frac{1}{m}dm\nonumber\\
&=&\frac{C\lambda}{\sigma\sqrt{2\pi}}\int^{m_{max}}_{m_{min}}\exp\left[-\frac{(\frac{t_o}{m})^2+\mu^2-2\mu\frac{t_o}{m}}{2\sigma^2}-\lambda m\right]\nonumber\\
&\times &\frac{1}{m}dm.
\label{prob_normal}
\end{eqnarray}
If the logarithm of the intrinsic timescales is normally distributed with mean $\tilde\mu$  and scale $\sigma_{sc}$ then:
\begin{equation}
P(t_i)=\frac{1}{t_i\sigma_{sc}\sqrt{2\pi}}\exp\left[-\frac{(\ln t_i-\tilde{\mu})^2}{2\sigma_{sc}^2}\right],
\label{intri_lognormal}
\end{equation}
The probability density function of observed timescales will be:
\begin{eqnarray}
&P(t_o)&=\frac{C\lambda \exp\left[-\frac{(\ln t_o-\tilde{\mu})^2}{2\sigma_{sc}^2}\right]}{t_o\sigma_{sc}\sqrt{2\pi}} \nonumber\\
&\times &\int^{m_{max}}_{m_{min}}\exp\left[-\frac{(\ln m)^2+2\ln m(\tilde{\mu}+\ln t_o)}{2\sigma_{sc}^2}-\lambda m\right]dm.\nonumber\\
\label{prob_lognormal}
\end{eqnarray}
By fitting equations \ref{prob_normal} and \ref{prob_lognormal} to the observed data we can optimize for  the mean ($\mu$) and standard deviation ($\sigma$) entering Eq. \ref{intri_normal} and the mean ($\tilde{\mu}$) and scale ($\sigma_{sc}$) entering Eq. \ref{intri_lognormal}.

The derivation of an analytical solution was only possible for the case of the $\delta$-function. The rest of the cases have to be solved either numerically or with the use of Monte-Carlo sampling.

\section{Assessing finite-sampling systematic effects}\label{system_effects}

\begin{figure}
\resizebox{\hsize}{!}{\includegraphics[scale=1]{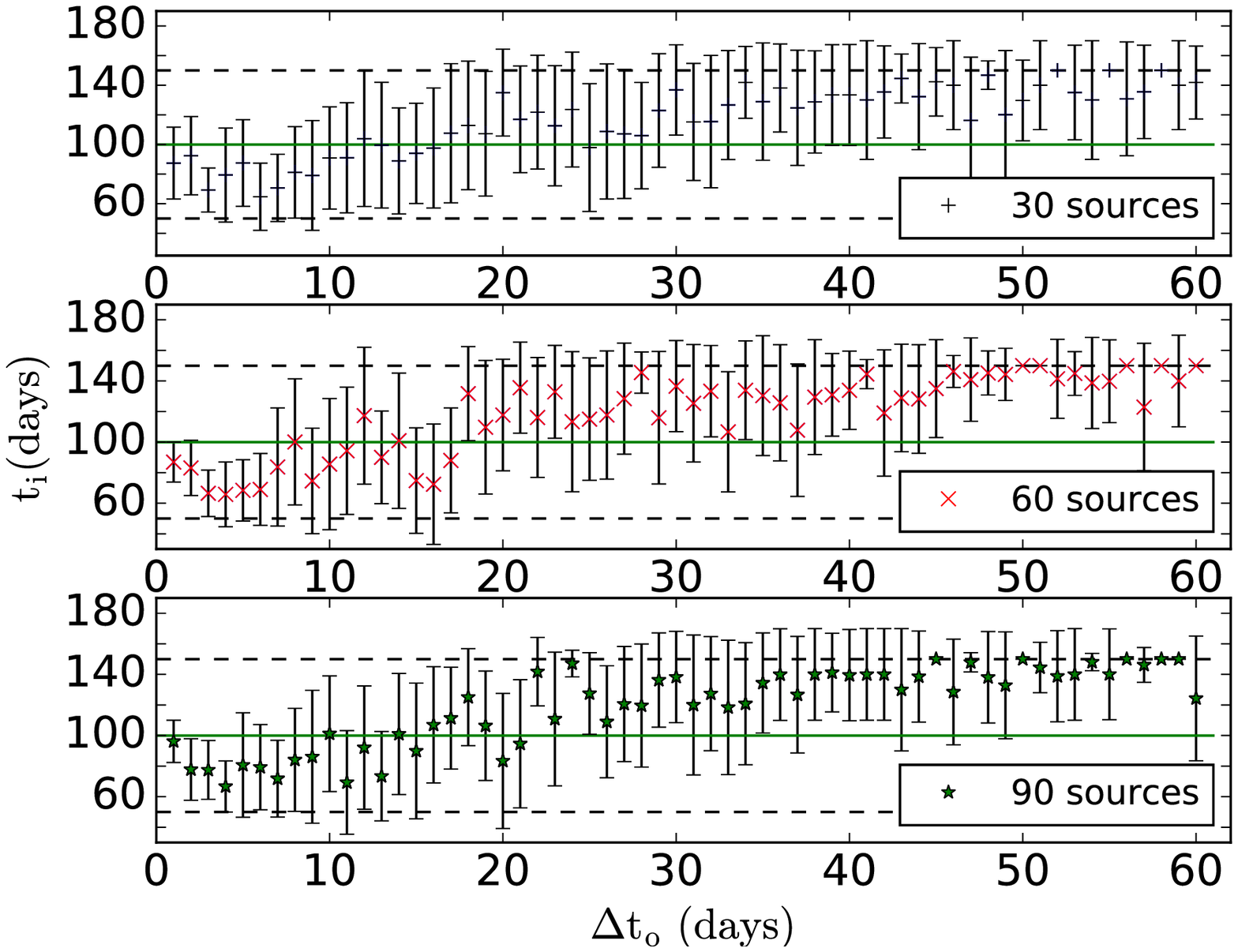} }
 \caption{Estimated best-fit $t_i$ versus interval of observations (both in days) for samples with pileup. The black ``+'' is for the 30 source sample, red ``x'' for the 60 source sample and the green ``$\star$'' for the 90 source sample. The error bars represent the uncertainty of the best-fit $t_i$. The green line at 100 days shows the position of the ``true'' intrinsic timescale, while the black dashed lines at 50 and 150 days show the limits of the parameter space.}
 \label{plt_time_vs_cad_sys}
 \end{figure}

As discussed in \S \ref{benchmarking}, the finite sampling of a light curve could lead to a systematic offset in the observed timescale distribution, especially in cases were we are interested in the fastest event observed in a source (e.g flares). The time duration of events shorter than the time interval between observations will be observed with duration equal to that time interval, thus creating pileups. For this reason, knowledge of the survey  $\Delta t_o$ is important in understanding such observational bias. A simple comparison between $\Delta t_o$ and the minimum observed timescale can provide insight on the existence (or not) of pileups in the data. The situation becomes even more complicated in highly variable sources  where it is possible to observe multiple overlapping flares within a short period of time. The blending of the flares will lead to overestimating their duration according to the survey $\Delta t_o$. In this case, a more sophisticated approach is required, one that will take into account such overlap, mitigating the effects of finite sampling \citep{Liodakis2016}.   

Here we focus only on the simple case where the fastest events observed are those with duration equal to the survey $\Delta t_o$. To examine the effects and systematic shifts induce by the sampling of our survey in this case, we repeat the procedure described in \S \ref{benchmarking}. This time, when creating the ``simulated-observed'' sample, timescales shorter than the time interval of observations are not discarded, but instead are set equal to that time interval. It is clear from Fig. \ref{plt_time_vs_cad_sys} that event pileup results in a significantly larger scatter in the estimated best-fit $t_i$ with a negative bias for small values of $\Delta t_o$ and a positive bias for larger values. There are also cases (samples produced with large $\Delta t_o$) where the estimated best-fit parameter is $t_i\approx 150$ (Fig. \ref{plt_time_vs_cad_sys}), which is the upper end of the parameter space. This means that these estimates have a $\gtrsim  50\%$ error. This effect is to be expected, since we are contaminating our simulated samples with longer timescales according to the chosen $\Delta t_o$. Moreover, the majority of cases have their 1$\sigma$ error reaching that level regardless of $\Delta t_o$. We conclude that the presence of pileup prevents the method from providing an accurate estimate and will lead to the overestimation of the intrinsic timescales.

Since many events in the time domain of blazar jets are connected to the size of the emission region through causality arguments, such an overestimation would also lead to the overestimation of that region's size and induce scatter or create artificial correlations between events in the time domain and physical properties of blazar jets. Thus datasets should be treated with great caution with respect to the $\Delta t_o$ of the survey in the application of the method.

\section{Assessing the effects of limited observing season length}\label{season_len}
\begin{figure}
\resizebox{\hsize}{!}{\includegraphics[scale=1]{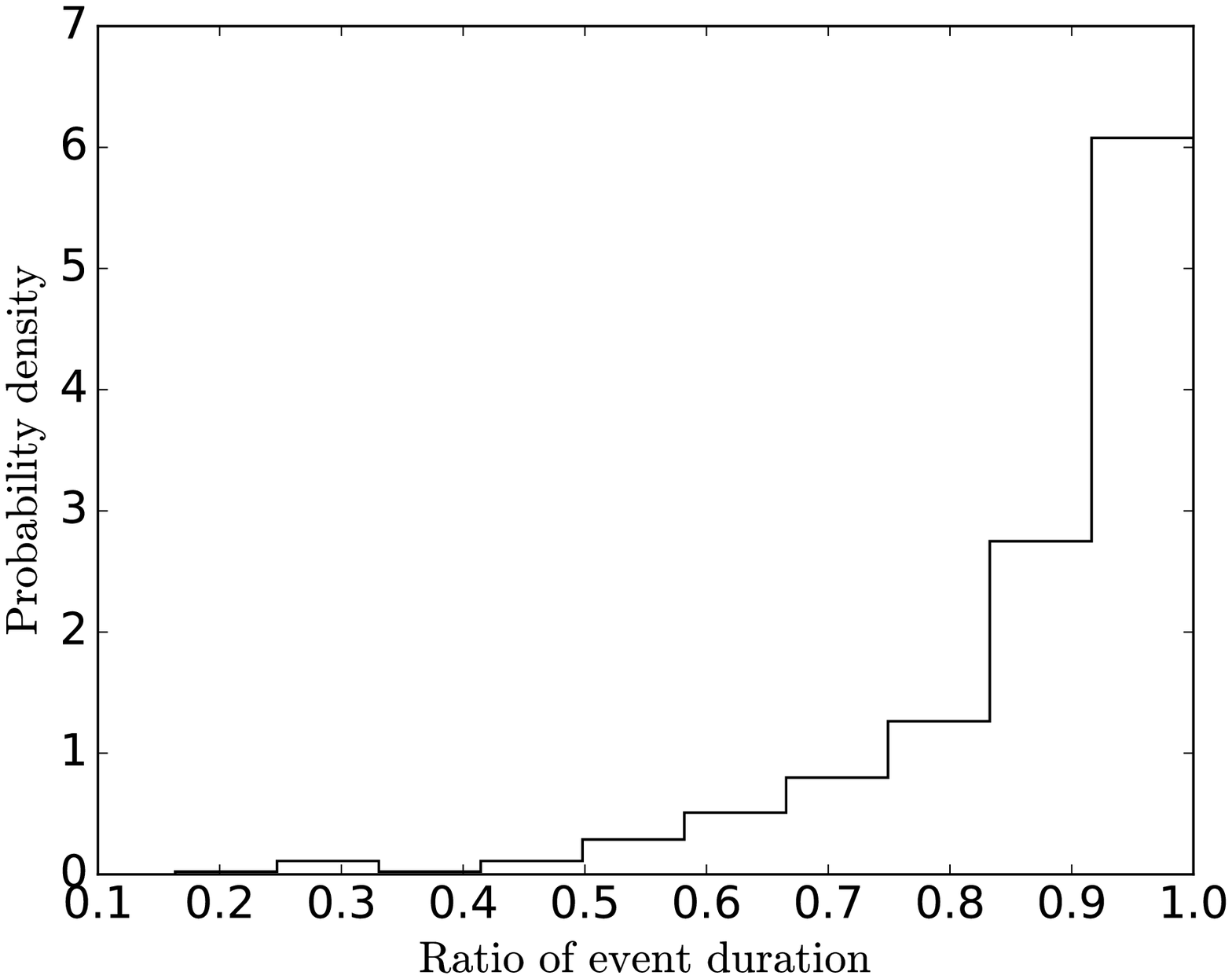} }
 \caption{Distribution of the ratio of durations ($T_i/T_{i+1}$) for all the EVPA events in each source observed with the definition adopted in this work for the RoboPol flux-limited subsample. }
 \label{plt_ratio_event}
 \end{figure} 
Our simulations show a surplus of long EVPA rotations (Fig. \ref{plt_robopol_normal_uniform}). As discussed in \S \ref{robopol_int_time} this is due to observational constrains related to the amount of time a source is available, with respect to the total RoboPol observing season length. Although the RoboPol sample was selected so that the sources would be available for the majority of the Skinakas observatory's observing period, it is not always the case. A limit on the observing season length sets an upper limit on the longest rotation the survey is able to detect, which in the best case scenario is the length of that observing season. Moreover, the time gap between observing seasons can affect the observed duration of any time-like event including an EVPA rotation. It is often unknown whether a rotation has begun prior to the beginning of the observing season, or if it continues after its end. Thus the observed time duration of an event can be significantly shorter than the ``true'' duration. 

Figure \ref{plt_robopol_normal_uniform} (upper panel) shows the observed distribution of EVPA rotations. There is a peak at $\sim$ 40 days after which the distribution rapidly declines for  longer events. Here we examine the possibility that the ``true'' observed distribution extends to 250 days (approximately the length of the RoboPol observing season) or longer, but the limited availability of the sources prevents us from observing longer rotations than the ones in Fig. \ref{plt_robopol_normal_uniform}. 

For every source, we sort the observed events by increasing duration,  calculate the ratios $T_i/T_{i+1}$, and construct the distribution of ratios (Fig. \ref{plt_ratio_event}) for the whole sample. This ratio is indicative of the spread of the observed durations between one event and the following longer/shorter event. Since each event in a given source is modulated by the same modulator factor, the ratio is independent of any relativistic effects, which allows us to combine all the ratios for individual sources in one distribution.

Using the above distribution, and assuming that the ``true'' duration of events is uniformly distributed from 3 (the shortest observed duration) to 250 days, we use random sampling to create a simulated dataset of the longest events that would have been observed given all the observational constrains (observing gap, limited availability, random event starting time with respect to beginning and end of the observing season) of the RoboPol survey. We then compare the distribution of simulated durations to the observed (Fig. \ref{plt_robopol_normal_uniform}) using the K-S test. We repeat the process $10^6$ times and calculate the number of trials for which the K-S test could not reject the null hypothesis that the samples are drawn from the same distribution (i.e. the probability value was $>$5\%). We find that $\sim 0.5\%$ of the trials resulted in a distribution consistent with the observed. If we extend the ``true'' duration of the events to 500 days, the number of trials drops to  $<10^{-6}$. Thus it is unlikely that the ``true'' observer's frame duration of EVPA events is much longer, yet due to  observational constrains we are not able to observe them. However, the fact that our method predicts the existence of events longer than what is observed in the data suggests that, although not very significant, there is a bias towards shorter in time events due to the availability of a source. Our findings stress the importance of long term uninterrupted observations in uncovering the true nature of EVPA rotations.

\label{lastpage}

\end{document}